\def\lsco{La$_{2-x}$Sr$_x$CuO$_4$}
\def\lbco{La$_{2-x}$Ba$_x$CuO$_4$}
\def\ybco{YBa$_2$Cu$_3$O$_{6+z}$}
\def\lsczo{La$_{2-x}$Sr$_x$Cu$_{1-y}$Zn$_{y}$O$_4$}

\documentclass[aps,prb,twocolumn,superscriptaddress]{revtex4}

\input epsf.sty

\begin{document}


%
%

\title{Magnetic properties of the overdoped superconductor
La$_{2-x}$Sr$_{x}$CuO$_{4}$ with and without Zn impurities}

\author{S. Wakimoto}
\email[Corresponding author: ]{swakimoto@neutrons.tokai.jaeri.go.jp}
\affiliation{ Department of Physics, University of Toronto, Toronto,
   Ontario, Canada M5S~1A7 }
\affiliation{ Advanced Science Research Center, 
   Japan Atomic Energy Research Institute, Tokai, 
   Ibaraki 319-1195, Japan. }

\author{R. J. Birgeneau}
\affiliation{ Department of Physics, University of Toronto, Toronto,
   Ontario, Canada M5S~1A7 }
\affiliation{ Department of Physics, University of California, 
   Berkeley, Berkeley, California 94720-7300 }

\author{A. Kagedan}
\affiliation{ Department of Physics, University of Toronto, Toronto,
   Ontario, Canada M5S~1A7 }

\author{Hyunkyung Kim}
\affiliation{ Department of Physics, University of Toronto, Toronto,
   Ontario, Canada M5S~1A7 }

\author{I. Swainson}
\affiliation{ Neutron Program for Materials Research, National
  Research Council of Canada, Chalk River, Ontario, Canada K0J~1J0 }

\author{K. Yamada}
\affiliation{ Institute of Material Research, Tohoku University, 
   Katahira, Sendai 980-8577, Japan }

\author{H. Zhang}
\affiliation{ Department of Physics, University of Toronto, Toronto,
   Ontario, Canada M5S~1A7 }

\date{\today}

\begin{abstract}

The magnetic properties of the Zn-substituted overdoped high-$T_c$
superconductor La$_{2-x}$Sr$_{x}$Cu$_{1-y}$Zn$_{y}$O$_{4}$ have been
studied by magnetization measurements and neutron scattering.
Magnetization measurements reveal that for Zn-free samples with $x
\geq 0.22$ a Curie term is induced in the temperature dependence of 
the magnetic susceptibility implying the existence of local 
paramagnetic moments. The induced Curie constant corresponds to 
a moment of 0.5 $\mu_B$ per additional Sr$^{2+}$ ion that exceeds 
$x=0.22$.  Zn-substitution in the overdoped \lsco~ also induces a 
Curie term that corresponds to 1.2 $\mu_B$ per Zn$^{2+}$ ion, 
simultaneously suppressing $T_c$.  The relationship between $T_c$ 
and the magnitude of the Curie term for Zn-free \lsco~ with 
$x \geq 0.22$ and for Zn-substituted \lsco~ with $x = 0.22$ are 
closely similar.  This signifies a general competitive relationship
between the superconductivity and the induced paramagnetic moment.
Neutron scattering measurements show that Zn-substitution
in overdoped \lsco~ anomalously enhances the inelastic magnetic
scattering spectra around the $(\pi, \pi)$ position, peaking at 
$\omega \sim 7$~meV.  These facts are discussed on the basis of 
a ``swiss-cheese" model of Zn-substituted systems as well as a 
microscopic phase separation scenario in the overdoped region 
indicated by muon-spin-relaxation measurements.

\end{abstract}

\pacs{74.72.Dn, 75.40.Gb, 75.40.Cx 61.12.Ex}

\maketitle

\section{Introduction}

Neutron scattering studies of the magnetic response in the hole-doped
high-$T_c$ cuprates up to the optimally doped region have revealed 
rich and complex behaviour of the static and dynamic spin 
correlations.  For the \lsco~ (LSCO) system, an incommensurate 
modulation at low energies has been found around the AF wave
vector~\cite{Yoshizawa_88,Bob_89,Cheong91} with the 
incommensurability, defined as the inverse of the modulation period, 
proportional to the hole concentration.~\cite{Yamada98}  Moreover, 
at the lower critical hole concentration of superconductivity, with 
increasing doping the modulation direction changes from the diagonal 
Cu-Cu direction in the insulator region to the parallel Cu-O-Cu 
direction in the superconducting (SC) 
region.~\cite{waki_rapid,waki_full,Matsuda_00,Fujita_02}  Such
incommensurate magnetic features have been discussed on the basis 
of a microscopically inhomogeneous electronic state of which the 
most commonly discussed example is 
charge stripes.~\cite{Machida_89,Schulz_89,Zaanen_89,Emery_97}

For the \ybco~ (YBCO) system, although the same type of low energy 
incommensurate magnetic response has been
confirmed,~\cite{DaiPRL98,Mook_Nature98,Arai_99,Dai01PRB,chirs03} 
the most prominent feature is a resonance peak that originates from 
a strong magnetic excitation at intermediate energies ($\sim 40$~meV) 
at the commensurate AF vector.  It  has been hypothesized that the 
resonance is directly related to 
the superconductivity.~\cite{Dai01PRB,chirs03,Regnault_95,Fong_00}
The resonance feature has been observed for other hole doped 
high-$T_c$ cuprates as well,~\cite{Fong_99,He_02} although until 
recently attempts to establish the universality of the resonance 
peak have been unsuccessful.  However, recent high energy measurements 
on YBCO,~\cite{Hayden_Nature_04,Chris_04,Reznik_03}
\lbco,~\cite{Tranquada_Nature_04} and LSCO~\cite{Christensen_04} have
indeed suggested a universal magnetic response; specifically, in each 
material the low energy incommensurate excitation disperses inwards 
with increasing energy and meets at the commensurate AF vector; the 
excitation then again disperses outwards in an approximately 
symmetric fashion in the high energy region.

These facts suggest that subtly different spectral weight
distributions in the universal dispersion discussed above give rise 
to the apparently different magnetic responses in each compound.  
In the case of the LSCO system, there is a relatively large spectral 
weight in the low energy incommensurate excitations.  The LSCO system 
provides the fortunate possibility of systematically studying the 
behaviour over an extremely wide range of hole concentrations from 
the underdoped to the overdoped region.  To-date, however, the 
magnetic properties in the overdoped region are not well understood.
Recent neutron scattering experiments~\cite{waki_04PRL} have revealed 
a surprising feature in the magnetism in overdoped LSCO.  These 
measurements show that the integrated dynamic spin susceptibility 
$\chi''(\omega)$ has a maximum at $\omega \sim 6$~meV and this 
maximum $\chi''(\omega)$ decreases linearly with $T_c$ for samples 
with $x \geq 0.25$.  Finally the low energy magnetic response 
disappears coincident with the disappearance of bulk 
superconductivity at $x = 0.30$.  Therefore, the low-energy 
incommensurate excitations are intimately connected with the 
superconductivity.

In the present paper we report a study of the magnetism in overdoped 
LSCO with and without Zn impurities.  Neutron scattering and 
magnetic susceptibility measurements elucidate the dependence of the 
superconductivity on certain features of the low energy magnetism in 
the overdoped region.
Magnetic susceptibility measurements of LSCO with $0.18 \leq x \leq
0.28$ reveal that the temperature dependence of the susceptibility
$\chi(T)$ for $x \leq 0.22$ falls on the universal curve originally 
reported by Johnston {\it et al.},~\cite{Johnston_PRL} while 
$\chi(T)$ for $x \geq 0.22$ exhibits in addition Curie paramagnetism;
these results are qualitatively consistent with those of Oda {\it et
al.}~\cite{Hokudai_1,Hokudai_2,Hokudai_3}  By comparing results in 
overdoped LSCO with and without Zn-impurities, it is found that both 
hole-overdoping and Zn-substitution induce Curie paramagnetism as 
well as a decrease in $T_c$ in the same manner, although 
Zn-substitution has a stronger effect.  This implies a general 
competitive relationship between the superconductivity and the 
${\rm \bf q} = 0$ Curie paramagnetism.
Neutron scattering experiments for Zn-substituted LSCO with $x=0.20$
and 0.25 exhibit an anomalous enhancement of the incommensurate 
inelastic magnetic scattering around the $(\pi ~\pi)$ position.  
Zn-substituted samples exhibit the same incommensurability and 
coherence length, while the enhancement in the $x=0.25$ sample is 
smaller than that in $x=0.20$.  These facts are apparently consistent 
with a ``swiss-cheese" model of Zn-impurity effects~\cite{Nachumi_96} 
and a microscopic phase separation model of the overdoped cuprates 
hypothesized from $\mu$SR measurements.~\cite{Uemura_03}

The organization of this paper is as follows.  In Sec. II, the 
experimental details are presented.  Results of the magnetic
susceptibility measurements of \lsczo~(LSCZO) together with an 
empirical scaling relation of the high temperature susceptibility are 
introduced in Sec. III.  Neutron scattering results of hole-overdoped 
LSCZO are reported in Sec. IV.  Finally, the results are discussed on 
the basis of the swiss-cheese model and the phase separation model in 
Sec. V and then summarized in Sec. VI.

\section{Experimental details}

For the magnetization measurements, a series of ceramic samples of 
LSCZO were synthesized by the conventional solid state reaction 
method.  The appropriate stoichiometric quantities of the starting
materials, La$_{2}$O$_{3}$, SrCO$_{3}$, CuO, and ZnO were ground and
baked in air for 12 hours at 900 $^{\circ}$C followed by an additional
grinding process.  The reacted powder was then pelletized under
hydrostatic pressure and sintered in air at 900 $^{\circ}$C for
10 hours and at 1150 $^{\circ}$C for an additional 10 hours.  
Finally, the pelletized samples were annealed in flowing oxygen gas 
at 850 $^{\circ}$C for 12 hours to restore any oxygen deficiencies in 
their composition.

The magnetic susceptibility measurements were carried out using a Quantum
Design physical property measurement system (PPMS).  The samples were
mounted at the end of a plastic straw with a small amount of Kapton
tape.  The sample chamber was purged 20 times at room temperature
before each measurement in order to minimize any air presence in
the measurement environment.  The SC shielding effect was measured
with a field of 10~Oe after cooling under zero field to determine
the SC transition temperature $T_c$.  The high temperature magnetic
susceptibility was measured with a field of $5$~T over a temperature
range of $20 K$ to $350 K$ after cooling under zero field.

For the neutron scattering measurements, single crystals of LSCZO  
were grown by the travelling solvent floating zone 
method.~\cite{Hosoya94,CHLee98}  Feed rods were prepared in the same
manner as the ceramic samples mentioned above aside from the addition 
of excess CuO of 2 to 3~mol\% to compensate for the evaporation of 
CuO during the high temperature growth.  The grown crystals were cut 
into the proper size for neutron scattering and then were annealed 
in flowing oxygen for 100 hours at 900 $^{\circ}$C.  Each prepared 
crystal had a typical size of 7 mm in diameter and 3.5 cm in length.

Inelastic neutron scattering experiments were performed using the C5
spectrometer at Chalk River Laboratory for the all samples except the 
Zn-free $x=0.20$ sample and the TAS1 spectrometer at the Japan Atomic 
Energy Research Institute for the Zn-free $x=0.20$ sample.  A fixed 
final energy ($E_f$) of 14.5~meV ($\lambda = 2.37$~\AA) and the 
collimation sequence of 33$'$-48$'$-51$'$-120$'$ were used for the 
former spectrometer while for the latter the configuration was $E_f = 
14.7$~meV and 40$'$-80$'$-80$'$-open.  Pyrolytic graphite (PG) 
crystals were used as a monochromator and an analyzer.  A PG filter 
was placed between the sample and the analyzer to eliminate higher 
order neutrons with wave lengths $\lambda/2$ and $\lambda/3$.
For each concentration, two single crystals were co-aligned on an Al 
sample holder with the $a$ and $b$ axes in the scattering plane, and 
then mounted in a closed cycle He refrigerator.  

All samples except the $x=0.20$ sample are tetragonal down to the 
lowest temperature with typical lattice constants of $a=b=3.73$~\AA,
corresponding to reciprocal lattice units of $a^{*}=b^{*}=
1.68$~\AA$^{-1}$ in tetragonal notation with space group $I4/mmm$.  
The $x=0.20$ sample exhibits a structural phase transition from the 
high-temperature tetragonal to the low-temperature orthorhombic 
phase at $\sim 70$~K, in good agreement with previous 
work.~\cite{Takagi}  Throughout this paper, we use tetragonal 
notation to express the indices.

\section{Magnetic susceptibility}

\subsection{Scaling behaviour}

Before we present the results of our magnetic susceptibility
measurements, we introduce the previously determined empirical 
scaling relation.  Johnston~\cite{Johnston_PRL} has systematically 
measured the temperature dependence of the magnetic susceptibility 
$\chi(x, T)$ of LSCO for $0 \leq x \leq 0.2$ and represented the 
data by the formula
\begin{equation}
\chi(x, T) =  \chi_{0}(x) + \chi^{2D}(x, T)
\end{equation}
where $\chi_{0}$ is the temperature independent uniform susceptibility
and $\chi^{2D}(x, T)$ is the temperature dependent in-plane 
susceptibility.  Johnston has reported that if one assumes the proper
value for $\chi_{0}(x)$, then $\chi^{2D}$ follows the simple scaling 
relation
\begin{equation}
F(T/T_{max}(x))
=\frac{\chi^{2D}(T,x)}{\chi_{max}^{2D}(T_{max},x)},
\end{equation}
where $F$ is a universal function and $\chi_{max}^{2D}$ is the 
maximum value of $\chi^{2D}$ at $T=T_{max}$.
This scaling relation has been re-examined in detail by Oda {\it et
al.}~\cite{Hokudai_1} for a wide concentration range of LSCO and LBCO.
They observed that beyond the hole concentration of $x=0.18$ in LSCO, 
$T_{max}$ saturates and $\chi(x, T)$ begins to deviate from the
universal scaling function.  Furthermore, this deviation takes the
form of a Curie paramagnetic term.  Thus, in the overdoped region,
the susceptibility can be expressed as
\begin{equation}
\chi(x, T) =  \chi_{0}(x) + \chi^{2D}(x, T) + \frac{C}{T}
\end{equation}
where $C$ is the Curie constant.

\begin{figure}
\centerline{\epsfxsize=3.3in\epsfbox{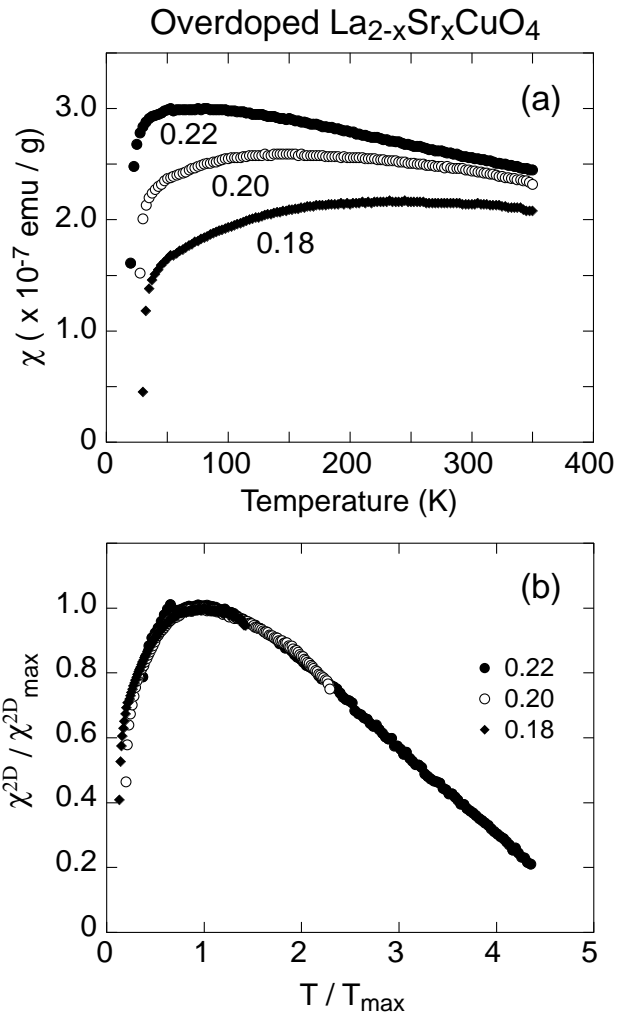}}
\caption{(a) Temperature dependence of the magnetic susceptibility for 
$x=0.18$, 0.20, and 0.22.  The extreme drop off in $\chi(T,x)$ at very
low $T$ is due to the SC transition.  (b) Universal scaling function,
$F(T/T_{max})=\chi^{2D}/\chi_{max}^{2D}$, for the noted hole
concentrations.}
\end{figure}

We present first the scaling relation in the current polycrystalline
samples that were annealed systematically in the same manner.  The 
temperature dependences of the magnetic susceptibility, $\chi(x, T)$, for
hole concentrations of $x=0.18$, 0.20, and 0.22 are shown in Fig 1
(a).  The data display a clear shift in $T_{max}$ to lower $T$ with
increasing hole concentration, qualitatively consistent with 
previous results.~\cite{Johnston_PRL,Hokudai_1,Hokudai_2}  These
$\chi(x, T)$ curves can be scaled excellently by Eq. (2) as shown in
Fig. 1 (b).  The values of $T_{max}$, $\chi_{max}$ and $\chi_{0}$ used 
to scale the data are summarized in Table 1, where $\chi_{max}$ is
defined as $\chi_{max} = \chi_{max}^{2D} + \chi_0$.  The increase of 
$\chi_0$ with increasing hole concentration $x$ is probably due to the
increase of the Pauli susceptibility, again qualitatively consistent 
with previous results.

\begin{table}
  \caption{$T_{max}$, $\chi_{max}$  and $\chi_{0}$ for hole
concentrations $x$ = 0.18, 0.20, 0.22.  $\chi_{max}$ is defined as
$\chi_{max} = \chi_{max}^{2D} + \chi_0$.}
\begin{ruledtabular}
\begin{tabular}{lccc}
$x$ & $T_{max}$~(K) & $\chi_{max}$~($10^{-7}$ emu/g) &
$\chi_{0}$~($10^{-7}$ emu/g) \\
\hline $0.18$ & $245.0$ & $2.167$ &
$0.5$ \\
$0.20$ & $152.7$ & $2.592$ &
$1.5$ \\
$0.22$ & $80.5$ & $2.99$ &
$2.3$ \\
\end{tabular}
\end{ruledtabular}
\end{table}

\subsection{Curie paramagnetism}

As the doping is increased beyond $x=0.22$, we have observed that a 
Curie paramagnetic term appears in the $\chi(x, T)$ curve.  Figure 2
(a) shows the $\chi(x, T)$ curves for $x=0.22$ and $0.24$, the latter
of which cannot be scaled by Eq. 2.  Instead, the differential
$\chi(x=0.24)-\chi(x=0.22)$ can be fitted excellently by the Curie 
term $C/T + \Delta\chi_0$ with Curie constant $C=1.25 \times 
10^{-6}$~emuK/g as
shown in Fig. 2 (b).  The constant term $\Delta\chi_0$ is included to
account for a possible change of the ``background'' susceptibility 
due to variations in both $\chi_0$ and the measurement environment.  
The excellent agreement between the data and the fit indicates that 
for $x > 0.22$ a simple Curie paramagnetic component appears in 
addition to the universally scaled $\chi^{2D}$.  Although there is a 
difference in the onset concentration where the Curie term starts to 
appear, possibly due to a difference in the post-annealing condition, 
the present results are qualitatively consistent with those of Oda 
{\it et al.}~\cite{Hokudai_1}  Thus the overdoping of charge carriers 
in the LSCO system induces Curie paramagnetism and reduces $T_c$.

\begin{figure}
\centerline{\epsfxsize=3.5in\epsfbox{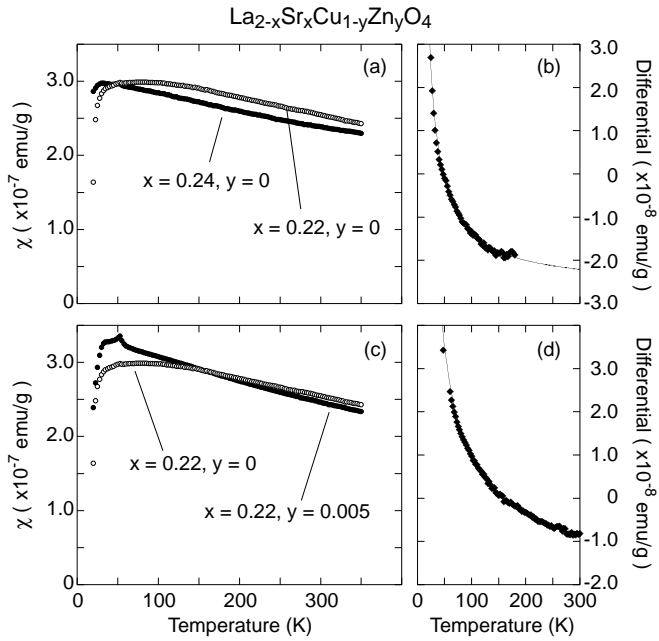}}
\caption{(a) Temperature dependence of the magnetic susceptibilities 
of Zn-free samples with $x=0.22$ and 0.24.  (b) Differential plot of
$\chi(x=0.24)-\chi(x=0.22)$.  (c) Magnetic susceptibilities of the 
Zn-free $x=0.22$ sample and a Zn-doped sample with $x=0.22$ and 
$y=0.005$.  A sharp small peak at $\sim 50$~K in the data for $(x, y)
=(0.22, 0.005)$ is due to the condensation of a small amount of oxygen in 
the sample chamber.  (d) Differential plot of $\chi(x=0.22, y=0)-
\chi(x=0.22, y=0.005)$.  For figures (b) and (d), the lines are the 
results of fits to the Curie term $C/T$.}
\end{figure}

\begin{figure}
\centerline{\epsfxsize=3.3in\epsfbox{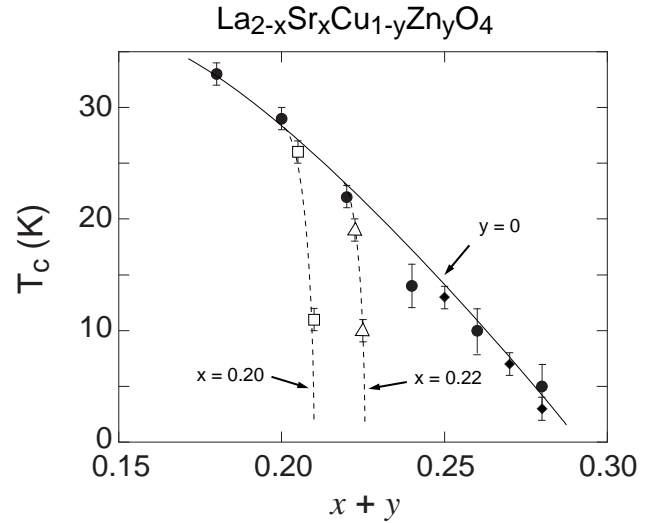}}
\caption{Decrease in $T_{c}$ due to the combination of increased hole 
concentration and Zn doping in the overdoped regime.  Note that the 
horizontal axis indicates the summation of both dopant concentrations 
$x+y$ of La$_{2-x}$Sr$_x$Cu$_{1-y}$Zn$_y$O$_4$.  The circles represent 
data for the Zn-free samples.  The diamonds represent $T_c$ for Zn-free 
single crystals referred from Ref.~\onlinecite{waki_04PRL}.  The 
squares and triangles represent data for the Zn-doped $x=0.20$ and 
$x=0.22$ samples, respectively.  The Zn amounts are $y=0.005$ and 0.01 
in order of decreasing $T_c$ for $x=0.20$, and $y=0.0025$ and 0.05 for 
$x=0.22$.}
\end{figure}

Similar to these effects, it is well known that the substitution
of small amounts of Cu$^{2+}$ ions with non-magnetic ions, such as
Zn$^{2+}$, induces Curie paramagnetism and eliminates
superconductivity.  To compare these effects, one by Sr-overdoping
and the other by Zn-substitution, we studied the magnetic
susceptibility of La$_{2-x}$Sr$_x$Cu$_{1-y}$Zn$_y$O$_4$.
Figure 3 highlights the correlation between the dopant concentration 
and the observed decrease in $T_c$ which implies the extinction of
superconductivity.  Note that the horizontal axis shows the total
amount of dopant $x+y$.  Closed circles, representing the data for
the Zn-free ($y=0$) samples, indicate a gradual decrease of $T_c$ with
increasing $x$.  $T_c$ for the Zn-free single crystals referred from
Ref.~\onlinecite{waki_04PRL} are also shown by diamonds, demonstrating 
excellent agreement with our results on the present powder samples.
In contrast, the data for the Zn-substituted samples with fixed Sr
concentrations $x=0.20$ and 0.22 indicated by squares and triangles,
respectively, show a drastic decrease of $T_c$ with increasing $y$.

\begin{figure}
\centerline{\epsfxsize=3.3in\epsfbox{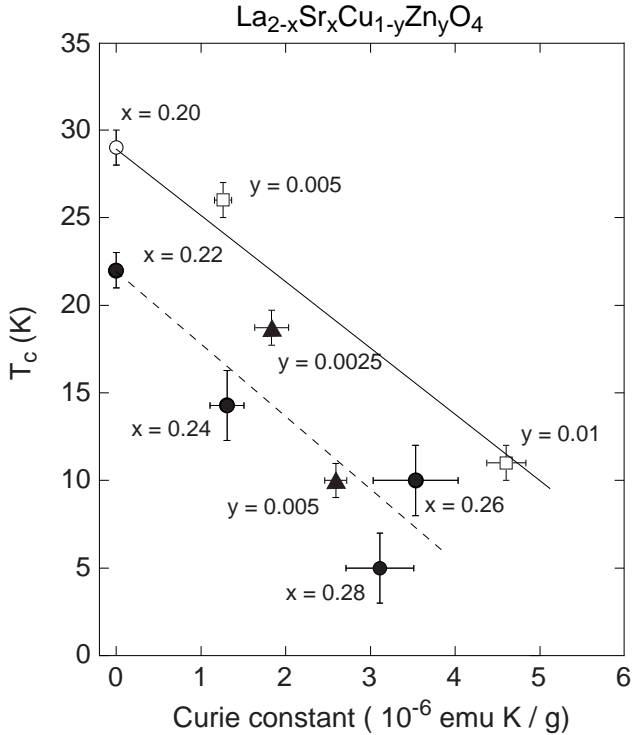}}
\caption{$T_c$ as a function of Curie constant for the La$_{2-
x}$Sr$_x$Cu$_{1-y}$Zn$_y$O$_4$ samples in the overdoped regime.
Circles are data for the Zn-free samples.  Squares and
triangles indicate data for Zn-doped $x=0.20$ and $x=0.22$ samples,
respectively.  The solid line is the result of a fit of the datum 
points for samples with $x=0.20$ shown by open symbols to a linear 
function $T_c = \alpha \times C + T_c(x=0.20, y=0)$, while the 
dashed line is the result of a fit of the rest of the datum points 
shown by closed symbols to a linear function $T_c = \alpha \times C + 
T_c(x=0.22, y=0)$.}
\end{figure}

As expected, Zn-substitution in the samples in the overdoped regime 
also induces Curie paramagnetism.  Figure 2 (c) shows $\chi(T)$ curves 
for two $x=0.22$ samples with $y=0$ and $y=0.005$.  The latter sample
shows a small sharp peak at $\sim 50$~K which arises from the
condensation of oxygen in the measurement environment as a result of 
inadvertently not achieving a proper vacuum in the measurement chamber 
for those particular data.
The differential of $\chi(x=0.22, y=0.005)-\chi(x=0.22, y=0)$ excluding
the temperature range where the oxygen peak appears is shown in Fig.
2 (d).  Again the differential can be fitted excellently by a simple 
Curie term with $C=2.59 \times 10^{-6}$~emuK/g.  Thus, Zn-substitution 
induces effects on the magnetism and superconductivity which are exactly
analogous to those caused by Sr-doping for $x \geq 0.22$.

The present data for the magnetic susceptibility for 
La$_{2-x}$Sr$_x$Cu$_{1-y}$Zn$_y$O$_4$ are summarized in Fig. 4. 
These data evince a clear correlation between a decrease of $T_{c}$ 
and an increase of the Curie constant.  From the results for the 
Zn-doped samples with fixed Sr concentration at $x=0.20$ shown by 
squares together with the datum point for the Zn-free sample indicated 
by an open circle, we conclude that increasing the Zn ion concentration 
reduces $T_c$ and increases the Curie constant.
The solid line is the result of a fit of these three datum points 
represented by open symbols to a 
linear function $T_c = \alpha \times C + T_c(x=0.20, y=0)$ which gives
$\alpha = -3.8(\pm 0.3) \times 10^{6}$~g/emu.  Similarly,
starting from a Zn-free $x=0.22$ sample, both further doping of Sr and 
Zn-substitution reduce $T_c$ as well as increasing $C$ in a similar 
manner as shown by the closed circles and triangles, respectively.  
These data may also be fit to a linear function 
$T_c = \alpha \times C + T_c(x=0.22, y=0)$.  The best fit, shown by 
the dashed line, is given by 
$\alpha = -4.2(\pm 0.6) \times 10^{6}$~g/emu, which is the same as 
that for the Zn-doped $x=0.20$ samples.
Thus, very interestingly, the overdoping of Sr for $x \geq 0.22$ and
Zn-substitution exhibit closely analogous effects, namely a decrease of 
$T_c$ with increasing $C$, although the Zn-substitution has
a much stronger effect.  This implies a general competitive relationship
between the superconductivity and the Curie paramagnetism.
Most importantly, it implies that in the overdoped regime for 
$x \geq 0.22$ increasing the Sr concentration creates local moments 
which in turn destroy the superconductivity.

\section{Dynamic antiferromagnetic correlations}

\begin{figure}
\centerline{\epsfxsize=3.3in\epsfbox{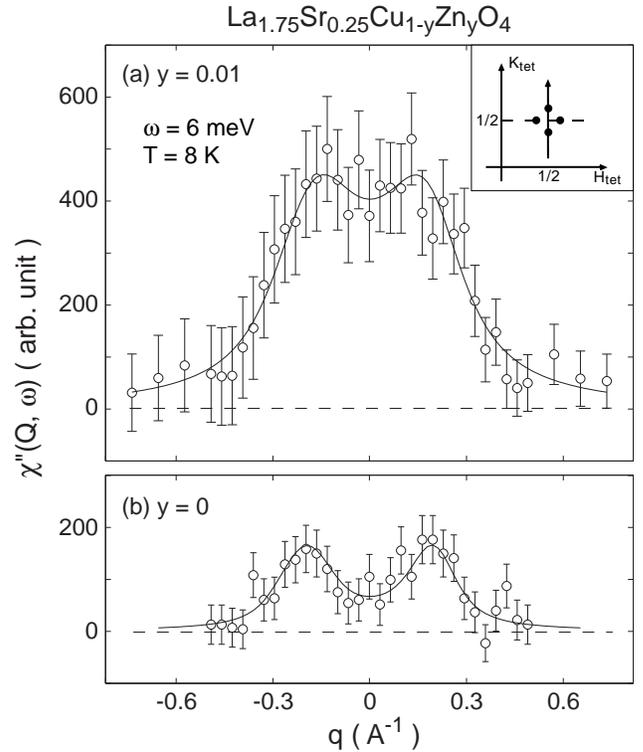}}
\caption{Dynamic spin susceptibility $\chi''({\rm \bf Q}, \omega)$ at
${\rm \bf Q}=(1/2+q, 1/2)$ and $\omega = 6$~meV for (a) 
Zn-substituted (y=0.01) and (b) Zn-free LSCO with $x=0.25$.
$\chi''({\rm \bf Q}, \omega)$ is calculated by normalizing to the
phonon intensity after a background subtraction.  The solid lines are 
the results of fits to a resolution-convoluted two-dimensional 
Lorentzian function.}
\end{figure}

Neutron scattering experiments have been performed to probe any
differences in the antiferromagnetic (AF) correlations between Zn-free
and Zn-substituted overdoped LSCO.
%
%
The observed magnetic cross-section can be written as
\begin{equation}
\frac{\partial^{2} \sigma}{\partial \Omega \partial \omega} =
   A \frac{1}{\hbar} p^{2} f^{2}(Q) e^{-2W}
   \frac{2 \chi''({\rm\bf Q}, \omega)}{\pi \mu_B^{2}} (n+1)
\end{equation}
where $p$ is the magnetic scattering length $0.27 \times 10^{-
12}$~cm, $f(Q)$ is the magnetic form factor, $e^{-2W}$ is the 
Debye-Waller factor, $\mu_B$ is the Bohr magneton, and $(n+1)$ is the
thermal population factor.  To compare the dynamic spin susceptibility
$\chi''({\rm \bf Q}, \omega)$ between Zn-free and Zn-substituted
samples, the observed cross-section has been normalized by
the integrated phonon intensity
\begin{equation}
I = A \frac{1}{2\omega_p} \frac{|{\rm\bf Q}|^2 \cos{\beta}}{M}
|F(Q)|^2 (n+1)
\end{equation}
where $\omega_p$ is the phonon frequency, $\beta$ is the angle
between {\bf Q} and the phonon polarization vector, $M$ is the 
molecular weight, and  $|F({\rm\bf Q})|$ is the structure factor.
Details of the normalization procedure are given in the Appendix 
including raw data of phonon and magnetic cross-sections.
%
The $\chi''({\rm \bf Q}, \omega)$ so obtained has been fitted to a
resolution-convoluted two-dimensional Lorentzian function to derive
the incommensurability $\delta$ and the intrinsic peak width 
$\kappa (\omega)$:
\begin{equation}
\chi''({\rm \bf Q}, \omega) \propto
    \sum_i \frac{1}{({\rm\bf Q}-{\rm\bf Q}_i)^{2} + \kappa (\omega)^{2}}
\end{equation}
where the summation over $i$ has been carried out for the four
incommensurate peaks at $(1/2 \pm \delta, 1/2)$ and $(1/2, 1/2 \pm
\delta)$.

Figure 5 shows $\chi''({\rm \bf Q}, \omega)$ for the Zn-free and 
Zn-substituted $x=0.25$ samples at $\omega = 6.2$~meV which is the 
energy at which the dynamic susceptibility of the Zn-free sample has 
a maximum.  The scan was made along the trajectory $(1/2, 1/2 + q)$ 
as shown by the arrow in the inset.  The data for $y=0$ in Fig. 5(b) 
are taken from Ref.~\onlinecite{waki_04PRL}.  Clearly, there is an 
anomalous enhancement of $\chi''({\rm \bf Q}, \omega)$ as a result 
of the Zn-substitution.
Furthermore, the Zn-substituted sample exhibits a somewhat flat-top 
shape profile possibly due to a smaller incommensurability and larger 
peak width.  Identical effects are observed in the $x=0.20$ samples
as shown in Fig. 6.   However, accounting for the different scales of 
the vertical axes in Fig. 6 and Fig. 5, it is evident that the 
$x=0.20$ sample with $y=0.005$ has a larger enhancement of 
$\chi''({\rm \bf Q}, \omega)$ than the $x=0.25$ sample with $y=0.01$.

\begin{figure}
\centerline{\epsfxsize=3.3in\epsfbox{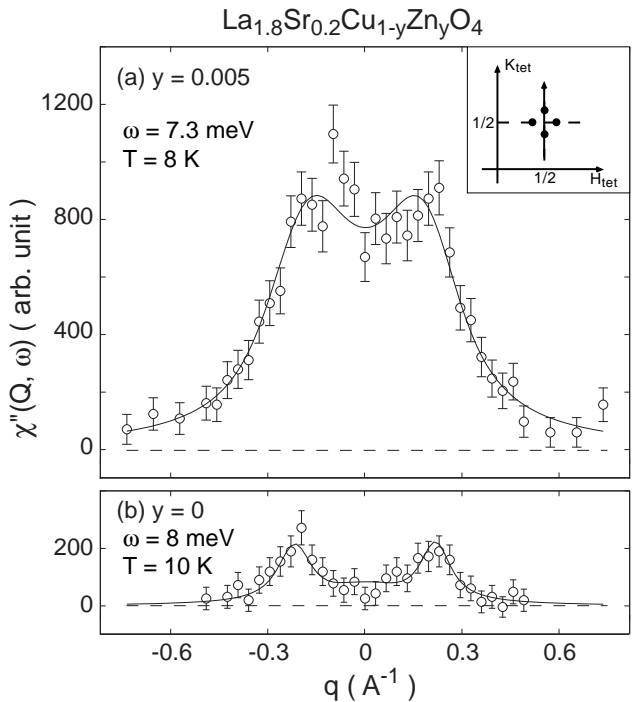}}
\caption{Analogous plots to Fig. 5 of $\chi''({\rm \bf Q}, \omega)$
for the (a) Zn-substituted (x=0.005) and (b) Zn-free LSCO with $x=
0.20$.  Note that the vertical axis scale is different from that of
Fig. 5.}
\end{figure}


\begin{table}
  \caption{Incommensurability $\delta$ in reciprocal lattice units
(r.l.u.) and intrinsic width $\kappa$
obtained by fitting the data to Eq. 6.}
\begin{ruledtabular}
\begin{tabular}{cccc}
$x$ & $y$ & $\delta$~(r.l.u.) & $\kappa (\omega)$~(\AA$^{-1}$) \\
\hline
$0.20$ & $0$ & $0.133(8)$ & $0.049(13)$ \\
$0.20$ & $0.005$ & $0.111(6)$ & $0.156(13)$ \\
$0.25$ & $0$ & $0.123(8)$ & $0.087(19)$ \\
$0.25$ & $0.01$ & $0.111(5)$ & $0.160(19)$ \\
\end{tabular}
\end{ruledtabular}
\end{table}

Before discussing the enhancement of $\chi''({\rm \bf Q}, \omega)$ in
detail, we first consider the changes in the incommensurability
$\delta$ and intrinsic width $\kappa (\omega)$ evinced by fitting the 
data to the resolution-convoluted Eq. 6.  The results of the fits are 
shown as the solid lines in Figs. 5 and 6, and the parameters $\delta$ 
and $\kappa (\omega)$ are summarized in Table 2.
Consistent with previous work,~\cite{Yamada98,CHLee00,CHLee03}
the incommensurability $\delta$ for the Zn-free samples saturates at
$\sim 1/8$.  However a small amount of Zn appears to reduce the 
incommensurability by $\sim 10$~\%.  The effect on $\kappa (\omega)$ 
is even more dramatic; the Zn-substituted samples show $\kappa (\omega)$ 
increased by a factors of 2 and 3 in the $x=0.25$ and $0.20$ samples, 
respectively.  It is interesting to note that the values of 
$\kappa (\omega)$ for the Zn-substituted samples are almost identical, 
although the Zn-free samples exhibit a very different values for 
$\kappa (\omega)$ for the samples with $x=0.20$ and $0.25$ respectively.

It is an open question why the incommensurability appears to
decrease in the Zn-doped samples.  We note that the peak 
profile for the sample with $x=0.25$ and $y=0.01$ in Fig. 5(a) can be 
alternatively reproduced approximately by a summation of a commensurate 
peak and the $\chi''({\rm \bf Q}, \omega)$ of the Zn-free $x=0.25$, 
implying the possibility that Zn-doping adds commensurate feature;
this is in place of the explanation given above that Zn-doping 
enhances the incommensurate magnetic correlations.  Either 
interpretation, however, 
is consistent with the conclusions drawn in the later sections.  

\begin{figure}
\centerline{\epsfxsize=3.3in\epsfbox{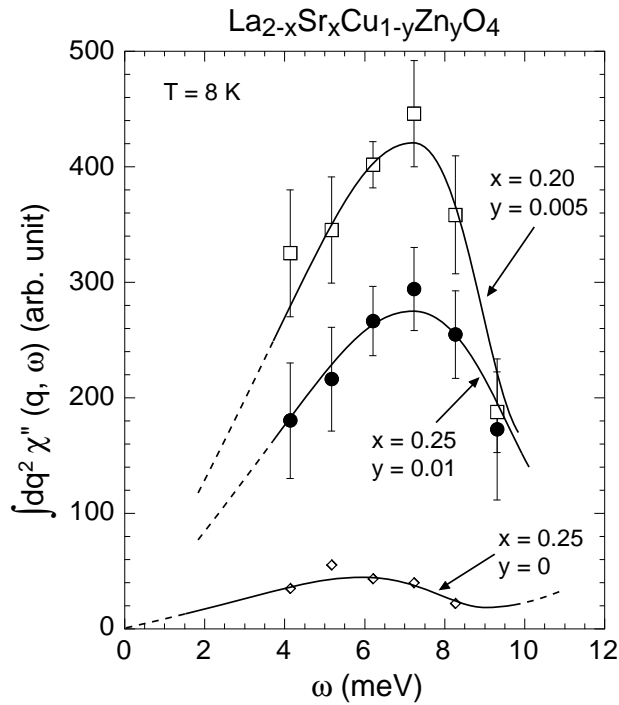}}
\caption{$\omega$-dependence of $q$-integrated dynamic spin
susceptibility of LSCZO samples together with the data for the 
Zn-free LSCO with x=0.25 referred from Ref.~\onlinecite{waki_04PRL}.
The lines are guides-to-the-eye.}
\end{figure}

Next, we present the enhancement of the spin susceptibility.  Figure
7 shows the $\omega$-dependence of the $q$-integrated dynamic spin
susceptibility $\int dq^{2} \chi''({\rm \bf q}, \omega)$ where the
integration is carried out over the four rods.  Corresponding to the
enhancement shown in Figs. 5 and 6, the Zn-substituted samples have
a large integrated $\chi''$.  
The Zn-free overdoped samples have been reported to have no clear 
spin gap down to $\omega = 2$~meV and show a linear increase with 
$\omega$.~\cite{waki_04PRL}  For the Zn-doped samples, no spin gap is 
observed down to $\omega=4$~meV and, moreover, no static magnetic 
peaks are observed.  At $\omega \leq 4$~meV, we presume a linear 
dependence with $\omega$ similar to that of the Zn-free samples 
as illustrated in Fig. 7.
Notably, the spectra for the samples doped with Zn retain their maxima 
at $\omega\sim 7$~meV which is very close to the energy where $\chi''$ 
of the Zn-free overdoped samples have their maxima.~\cite{waki_04PRL} 
This is in contrast to the case of Zn-substituted optimally doped LSCO 
$x=0.15$.~\cite{Kimura_03_PRL,Kofu_04}  In this case, while the $\chi''$
spectrum of Zn-free $x=0.15$ has a gap below $\sim 4$~meV and a
maximum at $\omega = 8$~meV, Zn-substitution gives rise to significant 
spectral weight inside the gap and with further substitution static 
($\omega = 0$) spin-density-wave (SDW) order appears.  This difference 
is discussed in detail in the next section.

\section{Discussion}

Our measurements of the uniform magnetic susceptibility, that is $q=0$
measurements, have revealed that a paramagnetic moment is induced by 
both Zn-substitution and by Sr-overdoping for $x \geq 0.22$ and in both 
cases the superconductivity is perturbed in the same manner.  
On the other hand, the neutron
scattering around $q=(\pi, \pi)$ demonstrates an anomalous enhancement
of the dynamic AF correlations as a result of Zn-substitution while the 
Zn-free sample shows a monotonic decrease in $\chi''$ with increasing Sr
concentration for $x \geq 0.25$.~\cite{waki_04PRL}  In this section, we
discuss the observed phenomena in detail to reconcile these observations.

\subsection{Paramagnetism and superconductivity}

The suppression of $T_c$ by magnetic perturbations has been discussed 
in pioneering work by Abrikosov and Gor'kov.~\cite{Abrikosov_61}  
Oda {\it et al.}~\cite{Hokudai_1} found that the decrease of $T_c$ as 
a function of the Curie constant $C$ for both overdoped LSCO and LBCO 
without Zn follows the Abrikosov-Gor'kov (A-G) theory.  In the present 
study, because of our limited amount of data, fits of the data in Fig. 
4 to the A-G theory itself yields results with unacceptably large 
uncertainties.  Hence we utilized a linear function which is a 
reasonable approximation when the magnetic impurity concentration is 
small.  The agreement of the slope $\alpha$ for both the Zn-substituted 
and Sr-overdoped systems shown in Fig. 4 provides the important insight 
that the $T_c$ suppression with increasing 
Sr-concentration in the overdoped region of LSCO is concomitant with 
the induced paramagnetic moments as in the case of Zn-substitution.

The Curie constant is related to the local moment by
\begin{equation}
C = \frac{N p_{eff}^{2} \mu_{B}^{2}}{3k_B}
\end{equation}
where $N$ is the number of induced paramagnetic spins and $p_{eff}$ is
the effective moment in units of $\mu_{B}$.  Using this
formula with the assumption that $N=N_{Zn}$ ($N_{Zn}$ is the number of 
Zn impurities), for the Zn-substituted samples of the present study one 
finds $p_{eff} = 1.2 (\pm 0.3)$/Zn.  This is in agreement with the 
previous result of $p_{eff} = 1$/Zn for the Sr underdoped regime of 
$0.05 \leq x \leq 0.15$.~\cite{Xiao_90}  This is consistent with the 
intuitive picture that a non-magnetic impurity breaks up a local AF pair 
thence yielding a single free spin.  Perhaps more surprisingly, 
Sr-overdoped samples without Zn give $p_{eff} = 0.5 (\pm 0.1)$/Sr
if we assume $N=N_{x}-N_{0.22}$, that is, based on the empirical results, 
$p_{eff}$ is assumed to be given by the concentration of Sr ions that 
exceed $x=0.22$. An alternative approach is to assume that the induced 
moment $p_{eff} = \sqrt{g^2 S (S+1)}$ with $S=1/2$ is always 1.  
In this case, the change of $p_{eff}$ in the Sr case would be absorbed 
by a change in the number of local moments $N$ given by 
$N=(N_{x}-N_{0.22})/4$.  This implies that $\sim 1/4$ of the Sr ions 
that exceed $x=0.22$ induce pair-breaking of the local AF pairs.

It is an open question how 1/4 of the overdoped Sr or holes create
the paramagnetic moment.  As a possible scenario, 1/4 of the overdoped 
Sr$^{2+}$ ions could substitute on the Cu sites instead of the La
sites and thus act as non-magnetic impurities.  Since the LSCO system 
tends to be oxygen-vacant in the overdoped regime, this might be
crystallographically possible even though Sr$^{2+}$ has a larger ionic 
radius than Cu$^{2+}$.  However, in our view this is not likely since the
effects on the dynamic AF correlation are fundamentally different between 
Zn-substitution and Sr-overdoping as we will discuss later.  Alternatively,
an origin based on the overdoped holes, rather than the doped ions
themselves, may be possible; for example, if 1/4 of overdoped holes
were to choose the Cu 3$d$ orbital rather than O 2$p$, this would result
in the elimination of the Cu local moment just as in the case of 
Zn-substitution.  Further study is necessary to obtain a microscopic 
understanding.

\subsection{Zn-impurity effects on AF correlations}

Based on their $\mu$SR measurements, Nachumi {\it et al.}~\cite{Nachumi_96} 
have suggested a ``swiss-cheese" model of the Zn impurity effects.  
In this model, the magnetically affected area is located around the 
impurity Zn ion and, therefore, the observed enhancement of the spectral 
weight in $\chi''$ must originate from this area.  As shown in Table II, 
$\kappa (\omega)$ for the Zn substituted samples exhibits the same value 
$\sim 0.16$~\AA$^{-1}$ in spite of the different $\kappa (\omega)$ values 
of the Zn-free samples.  Correspondingly, the affected area has a length 
scale of $1/\kappa (\omega) \simeq 6$~\AA~ in the overdoped case which is 
quite a bit smaller than in the case of the optimally and underdoped LSCO.

As shown in Fig. 7, $\chi''$, which is greatly enhanced by a small amount 
of Zn substitution, nevertheless still peaks at $\omega = 7$~meV.  
This is qualitatively different from the behaviour in the optimally 
doped LSCO case. As revealed by neutron scattering,
a small amount of Zn-impurity ($y \leq 0.01$) in optimally doped LSCO
induces spectral weight in $\chi''$ in the spin-gap energy region and,
furthermore, spin-density-wave order develops with further Zn-substitution 
($y=0.017$).~\cite{Kimura_03_PRL,Kofu_04}  Our results
for the {\it overdoped} LSCO with Zn do not show either detectable
SDW order or a spectral peak shift to lower energies in comparison
to the Zn-free samples.  We speculate that the overdoped sample has a
characteristic spin fluctuation frequency that is higher than that in
the optimally doped sample and therefore, the induced $\chi''$ peaks at
a higher $\omega$ in the overdoped case than in the optimal case.  This 
may also be the reason why no static SDW state develops in overdoped
LSCO substituted with Zn as shown by the the absence of a wipeout effect 
observed by nuclear quadrupole resonance~\cite{Yamagata_03_NQR} and the 
absence of static magnetic order observed by muon-spin-relaxation
($\mu$SR).~\cite{Panagopoulos_04}

Another important insight into the nature of the overdoped cuprates has 
been suggested by $\mu$SR; the muon spin relaxation rate $\sigma$
which is proportional to the SC carrier density decreases with
increasing doping in overdoped Tl$_2$Ba$_2$CuO$_{6+
\delta}$.~\cite{Uemura_93,Niedermayer93}  This has been attributed to
microscopic phase separation of the charge carriers into local 
Cooper-paired SC states and unpaired Fermi liquid (FL) states.  The 
decreasing $T_c$ in the overdoped region is correlated with a 
corresponding decrease in the SC volume fraction.~\cite{Uemura_03}  
This scenario can be reconciled with our neutron scattering observations 
if we assume that the SC regions retain dynamic AF spin correlations 
that support Cooper pair formation while the FL region has no AF 
correlations and concomitantly no superconductivity.

In this scenario, the linear decrease of $\chi''$ with $T_c$ observed in
Ref.~\onlinecite{waki_04PRL} can be naturally understood as originating 
from the decrease of the SC volume fraction.  The complete disappearance
of $\chi''$ in the non-superconducting $x=0.30$ sample signifies the
100~\% volume fraction of the FL region.
The microscopic phase separation scenario also predicts a critical 
threshold of the hole concentration for the phase separation, implying 
a constant hole concentration in the SC regions in the overdoped
samples.  Our observations exhibit related effects on the AF 
correlations by Zn impurities in both the $x = 0.20$ and $0.25$ samples.  
Specifically, the induced magnetic peaks have the same $\kappa (\omega)$ 
and $\delta$, and furthermore the induced $\chi''$ has a maximum at the 
same energy, $\omega = 7$~meV, in both samples.  The only difference is 
that the enhancement of $\chi''$ is smaller for $x=0.25$ than for 
$x=0.20$.  Since the induced AF correlations originate from the SC 
region, the constancy of $\kappa (\omega)$ and $\delta$ is consistent 
with the constant hole concentration in the SC regions.  Furthermore, 
the smaller enhancement of $\chi''$ in the $x=0.25$ sample compared 
with that in the $x=0.20$ sample can be attributed to the smaller SC 
volume fraction in the $x=0.25$ sample.  
Consistent with this, our preliminary measurement
shows that Zn-substitution into a $x=0.30$ sample which has no SC
fraction shows no evidence for the inducement of observable 
AF correlations.

\section{Concluding remarks}

We have measured the uniform magnetic susceptibility and neutron
scattering of overdoped LSCO with and without Zn impurities.
Uniform susceptibility measurements show clearly that the
superconductivity in the overdoped region is perturbed by the induced 
Curie paramagnetism.  On the other hand, our neutron scattering results
are consistent with a microscopic phase separation scenario into local 
superconducting and Fermi liquid regions.  To reconcile these facts,
it is natural to speculate that the paramagnetic Curie behaviour 
originates from the non-superconducting Fermi liquid region in the 
Zn-free samples.  It is important to elucidate microscopically how 
the overdoped Sr ions or associated holes create the paramagnetic moments and 
the Fermi-liquid regions simultaneously.

As for the Zn-doped LSCO for $x \geq 0.20$, we have observed a 
surprisingly large enhancement of the AF correlations around 
$(\pi ~\pi)$ by a tiny amount of Zn impurity.  The enhancement in the 
overdoped samples far exceeds that in the optimally doped 
LSCO.~\cite{Kimura_03_PRL,Kofu_04}  As we discussed, our observations 
appear to be qualitatively consistent with the swiss-cheese model that 
limits the Zn impurity effect on the magnetism to the area around the Zn 
ions.  It is, however, disputable quantitatively that a small area around 
the Zn-impurities causes such a huge effect.  To elucidate the origin 
of the enhancement, it is also important to study the Zn-effects on the 
spin excitations in the high energy region.  

Finally, our measurements have evinced remarkable changes in the magnetic 
properties from the optimally doped to the overdoped regime: 
appearance of the Curie paramagnetism and anomalous enhancement of AF 
correlations by Zn impurities.  Moreover, around the lower boundary of 
the overdoped region, the conductivity changes from unusual-metallic 
to normal-metallic behaviour; other measurements suggest that the 
pseudogap closes in this same crossover region.  These facts 
may imply an intriguing phase crossover between the optimally- and over-doped 
regions around $x=0.20$ which is crucial to the superconductivity.  
Further theoretical and experimental studies of this crossover should give new 
insights into the physics of high-$T_c$ superconductivity.

\begin{acknowledgments}

The authors thank Z. Tun, J. M. Tranquada, C. Stock, G. Shirane, M.
Matsuda, B. Khaykovich, K. Kakurai, Y. Endoh, and W. J. L. Buyers for 
invaluable discussions.  Work at the University of Toronto is part of 
the Canadian Institute of Advanced Research and supported by the 
Natural Science and Engineering Research Council of Canada, while 
research at Tohoku University is supported by a Grant-in-Aid from the 
Japanese Ministry of Education, Culture, Sports, Science and 
Technology.

\end{acknowledgments}

\appendix*
\section{}

\begin{figure}
\centerline{\epsfxsize=3.3in\epsfbox{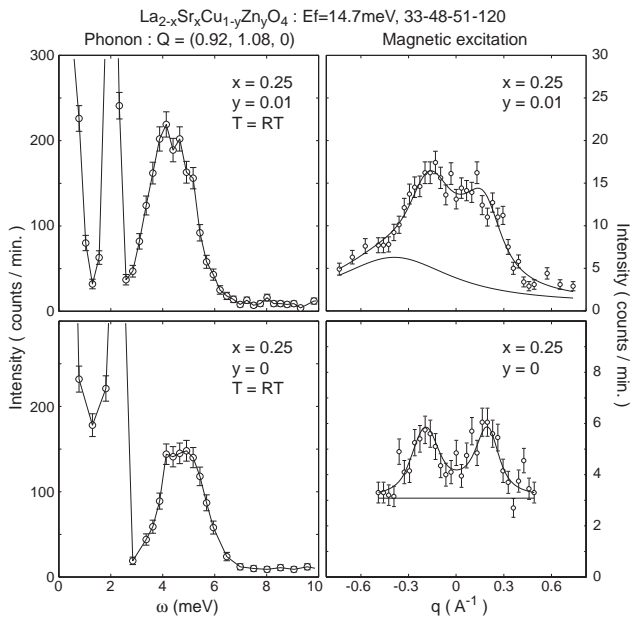}}
\caption{Raw data of phonon (left) and magnetic peaks (right) for the 
$x=0.25$ samples without and with Zn.  All profiles are measured at the 
C5 spectrometer with fixed $E_f = 14.5$~meV and a collimation sequence 
of 33$'$-48$'$-51$'$-120$'$.  The magnetic cross-section are collected 
by counting for 10 min./point for the Zn-doped and 20 min./point for 
the Zn-free samples.}
\end{figure}

To compare the magnetic profiles of the different samples directly, the
normalization has been made based on phonon intensities using Eqs.(4) and (5).  
The left side figures in Fig. 8 show raw data of phonon peaks of the 
$x=0.25$ samples with and without Zn measured at ${\rm \bf Q} = (0.92, 
1.08, 0)$, using the same spectrometer configuration.  
The peak at $\sim 2$~meV adjacent to the phonon peak at $\sim 4.5$~meV 
originates from the 
tail of the nuclear Bragg peak detected by the edge of the resolution ellipsoid, 
proving that scans are made on the focusing side.  Since the data are taken 
by fixing the final energy $E_f$, first the data have been corrected for 
higher order neutrons monitor rates at each incident energy $E_i$.  
Those for the Chalk River Reactor have been measured and reported in 
Ref.~\onlinecite{chirs03}, while those for JAERI reactor have been 
determined by a general formula in Ref.~\onlinecite{Gen_text}.  Then 
the phonon peaks have been fitted by a Gaussian line-shape to evaluate the 
phonon integrated intensity $I$ in Eq. (5).  Thus, one can now evaluate 
the parameter $A$.

Using the $A$ obtained above in Eq. (4), the observed magnetic cross-section 
may be directly connected to $\chi''({\rm \bf Q}, \omega)$ as normalized.  For this 
purpose, we estimated the net magnetic intensity by subtracting the background.  
The figures on the right in Fig. 8 show the observed raw profiles of the magnetic 
peaks.  While the figures are shown as the intensity for 1 min. counting 
time, the actual measurements were done by counting for 10 min. for the 
Zn-doped sample and for 20 min. for the Zn-free sample.  In analyses of 
the most corrected profiles, we have used a flat or sloping background.  
For the profile of the Zn-doped $x=0.25$ sample shown in the right-top figure, 
however, we have utilized a broadly peaked background since the 
wide-ranged profile naturally implies such a background.  Unfortunately, 
we were not able to determine the origin of the peaked background; however, 
assuming either a sloped or peaked background does not affect the results 
dramatically.  The background-subtracted magnetic intensities have been 
corrected for higher order neutron monitor rates and then the obtained 
cross-section have been determined as 
$\frac{\partial^{2} \sigma}{\partial \Omega \partial \omega}$ in 
Eq. (4).  Finally, the normalized $\chi''({\rm \bf Q}, \omega)$ shown in 
Fig. 5 are obtained from the raw data shown in Fig. 8.



\begin{references}

\bibitem{Yoshizawa_88} H. Yoshizawa, S. Mitsuda, H. Kitazawa and
K. Katsumata,
J.\ Phys.\ Soc.\ Jpn. {\bf 57}, 3686 (1988).

\bibitem{Bob_89} R.J. Birgeneau, Y. Endoh, K. Kakurai, Y. Hidaka, 
T. Murakami, M.A. Kastner, T.R. Thurston, G. Shirane, and K. Yamada,
Phys.\ Rev.\ B {\bf 39}, R2868 (1989).

\bibitem{Cheong91} S.-W. Cheong, G. Aeppli, T. E. Mason, H. Mook, S.
M. Hayden, P. C. Canfield, Z. Fisk, K. N. Clausen, and J. L. Martinez,
Phys. Rev. Lett. {\bf 67}, 1791 (1991).

\bibitem{Yamada98} K. Yamada, C. H. Lee, K. Kurahashi, J. Wada, S. Wakimoto,
S. Ueki, H. Kimura, Y. Endoh, S. Hosoya, G. Shirane, R. J. Birgeneau,
M. Greven, M. A. Kastner, and Y. J. Kim,
Phys. Rev. B {\bf 57}, 6165 (1998).

\bibitem{waki_rapid} S. Wakimoto, G. Shirane, Y. Endoh, K. Hirota, S.
Ueki, K. Yamada, R.J. Birgeneau, M.A. Kastner, Y.S. Lee, P.M. Gehring,
and S.H. Lee,
Phys. Rev. B {\bf 60}, R769 (1999).

\bibitem{waki_full} S. Wakimoto, R. J. Birgeneau, M. A. Kastner, Y.
S. Lee, R. Erwin, P. M. Gehring, S. H. Lee, M. Fujita, K. Yamada, Y.
Endoh, K. Hirota, and G. Shirane,
Phys. Rev. B {\bf 61}, 3699 (2000).

\bibitem{Matsuda_00} M. Matsuda, M. Fujita, K. Yamada, R. J.
Birgeneau, M. A. Kastner, H. Hiraka, Y. Endoh, S. Wakimoto, and G.
Shirane,
Phys. Rev. B. {\bf 62}, 9148 (2000).

\bibitem{Fujita_02} M. Fujita, K. Yamada, H. Hiraka, P. M. Gehring,
S.-H. Lee, S. Wakimoto, and G. Shirane,
Phys. Rev. B {\bf 65}, 064505 (2002).



\bibitem{Machida_89} K. Machida, Physica C {\bf 158}, 192 (1989).

\bibitem{Schulz_89} H. J. Schulz, J. Phys. France {\bf 50}, 2833 (1989).

\bibitem{Zaanen_89} J. Zaanen and O. Gunnarsson, Phys. Rev. B {\bf 40}, R7391 (1989).

\bibitem{Emery_97} V. J. Emery, S. A. Kivelson, and O. Zachar,
Phys. Rev. B {\bf 56}, 6120 (1997).





\bibitem{DaiPRL98} P. Dai, H. A. Mook, and F. Dogan,
Phys. Rev. Lett. {\bf 80}, 1738 (1998).

\bibitem{Mook_Nature98}
H. A. Mook, P. Dai, S. M. Hayden, G. Aeppli, T. G. Perring, and F.
Dogan,
Nature {\bf 395}, 580 (1998).

\bibitem{Arai_99} M. Arai, T. Nishijima, Y. Endoh, T. Egami, S.
Tajima,
K. Tomimoto, Y. Shiohara, M. Takahashi, A. Garrett, and S. M.
Bennington,
Phys. Rev. Lett. {\bf 83}, 608 (1999).

\bibitem{Dai01PRB} P. Dai, H. A. Mook, R. D. Hunt, and F. Dogan,
Phys. Rev. B {\bf 63}, 054525 (2001).

\bibitem{chirs03} C. Stock, W. J. L. Buyers, R. Liang, D. Peets, Z.
Tun, D. Bonn, W. N. Hardy, and R. J. Birgeneau
Phys. Rev. B {\bf 69}, 014502 (2004).

\bibitem{Regnault_95} L. P. Regnault, P. Bourges, P. Burlet, J. Y.
Henry, J. Rossa-Mignod, Y. Sidis, and C. Vettier,
Physica B {\bf 213\&214}, 48 (1995).

\bibitem{Fong_00} H. F. Fong, P. Bourges, Y. Sidis, L. P. Regnault,
J. Bossy, A. Ivanov, D. L. Milius, I. A. Aksay, and B. Keimer,
Phys. Rev. B {\bf 61}, 14773 (2000).

\bibitem{Fong_99} H. F. Fong, P. Bourges, Y. Sidis, L. P. Legnault,
A. Ivanov, G. D. Gu, N. Koshizuka, and B. Keimer,
Nature {\bf 398}, 588 (1999).

\bibitem{He_02} H. He, P. Bourges, Y. Sidis, C. Ulrich, L. P.
Regnault, S. Pailhes, N. S. Berzigiarova, N. N. Kolsenikov, and B.
Keimer,
Science {\bf 295}, 1045 (2002).

\bibitem{Hayden_Nature_04} S. M. Hayden, H. A. Mook, P. Dai,
T. G. Perring, and F. Dogan
Nature {\bf 429}, 531 (2004).

\bibitem{Chris_04}  C. Stock, W. J. L. Buyers, R. A. Cowley, P. S.
Clegg, R. Coldea, C. D. Frost, R. Liang, D. Peets, D. Bonn, W. N.
Hardy, and R. J. Birgeneau,
Phys. Rev. B {\bf 71}, 024522 (2005).

\bibitem{Reznik_03}  D. Reznik, P.Bourges, L. Pintschovius, Y. Endoh,
Y. Sidis, T. Masui, and S. Tajima,
Phys. Rev. Lett. {\bf 93}, 207003 (2004).

\bibitem{Tranquada_Nature_04} J. M. Tranquada, H. Woo, T. G. Perring,
H. Goka, G. D. Gu, G. Xu, M. Fujita, and K. Yamada,
Nature {\bf 429}, 534 (2004).

\bibitem{Christensen_04}  N. B. Christensen, D. F. McMorrow, H. M.
Ronnow, B. Lake, S. M. Hayden, G. Aeppli, T. G. Perring, M.
Mangkorntong, M. Nohara, and H. Tagaki,
Phys. Rev. Lett. {\bf 93}, 147002 (2004).

\bibitem{waki_04PRL} S. Wakimoto, H. Zhang, K. Yamada, I. Swainson,
Hyunkyung Kim, and R. J. Birgeneau,
Phys. Rev. Lett. {\bf 92}, 217004 (2004)

\bibitem{Johnston_PRL} D. C. Johnston,
Phys. Rev. Lett. {\bf 62}, 957 (1989).

\bibitem{Hokudai_1} M. Oda, T. Nakano, Y. Kamada, and M. Ido,
Physica C {\bf 183}, 234 (1991).

\bibitem{Hokudai_2} T. Nakano, M.Oda, C. Manabe, N. Momono, Y. Miura,
and M. Ido,
Phys. Rev. B {\bf 49}, 16000 (1994).

\bibitem{Hokudai_3} T. Nakano, N. Momono, T. Nagata, M. Oda, and M.
Ido,
Phys. Rev. B {\bf 58}, 5831 (1998).

\bibitem{Nachumi_96} B. Nachumi, A. Keren, K. Kojima, M. Larkin, G.
M.
Luke, J. Merrin, O. Tchernysh\"ov, Y. J. Uemura, N. Ichikawa, M.
Goto,
and S. Uchida,
Phys. Rev. Lett. {\bf 77}, 5421 (1996).

\bibitem{Uemura_03} Y. J. Uemura,
Solid State Comm. {\bf 126}, 23 (2003).

\bibitem{Hosoya94} S. Hosoya, C. H. Lee, S. Wakimoto, K. Yamada, and
Y. Endoh,
Physica C {\bf 235-240}, 547 (1994).

\bibitem{CHLee98} C. H. Lee, N. Kaneko, S. Hosoya, K. Kurahashi, S.
Wakimoto, K. Yamada, and Y. Endoh,
Semicond. Sci. Technol. {\bf 11}, 981 (1998).

\bibitem{Takagi} H. Takagi, R. J. Cava, M. Marezio, B. Batlogg, J. J.
Krajewski, and W. F. Peck, Jr., P. Bordet, and D. E. Cox,
Phys. Rev. Lett. {\bf 68}, 3777 (1992).

\bibitem{CHLee00} C. H. Lee, K. Yamada, Y. Endoh, G. Shirane, R. J.
Birgeneau, M. A. Kastner, M. Greven, and Y.-J. Kim, J.
Phys. Soc. Jpn. {\bf 69}, 1170 (2000);

\bibitem{CHLee03} C. H. Lee, K. Yamada, H. Hiraka, C. R. Venkateswara
Rao, and Y. Endoh,
Phys. Rev. B {\bf 67}, 134521 (2003).

\bibitem{Kimura_03_PRL} H. Kimura, M. Kofu, Y. Matsumoto, and K.
Hirota,
Phys. Rev. Lett. {\bf 91}, 067002 (2003).

\bibitem{Kofu_04} M. Kofu, H. Kimura, and K. Hirota,
Phys. Rev. B {\bf 72}, 064502 (2005).

\bibitem{Abrikosov_61} A. A. Abrikosov and L. P. Gor'kov,
Soviet Phys. JETP {\bf 12}, 1243 (1961);
M. Tinkham,
{\it Introduction to superconductivity} (McGraw-Hill, New York, 1996).

\bibitem{Xiao_90} G. Xiao, M. Z. Cieplak, C. L. Chien,
Phys. Rev. B {\bf 42}, 240 (1990).

\bibitem{Yamagata_03_NQR} H. Yamagata, H. Miyamoto, K. Nakamura, M.
Matsumura, and Y. Itoh,
J. Phys. Soc. Jpn. {\bf 72}, 1768 (2003).

\bibitem{Panagopoulos_04}  C. Panagopoulos, A. P. Petrovic, A. D.
Hillier, J. L.Tallon, C. A. Scott, and B. D. Rainford,
Phys. Rev. B 69, 144510 (2004).

\bibitem{Uemura_93} Y. J. Uemura, A. Karen, L. P. Le, G. M. Luke, W.
D. Wu, Y. Kubo, T. Manako, Y. Shimakawa, M. Subramanian, J. L. Cobb,
and J. T. Market,
Nature {\bf 365}, 805 (1993).

\bibitem{Niedermayer93} Ch. Niedermayer, C. Bernhard, U. Binninger,
H. Gl\"uckler, J. L. Tallon, E. J. Ansaldo, J. I. Budnick,
Phys. Rev. Lett. {\bf 71}, 1764 (1993).

\bibitem{Gen_text} G. Shirane, S. M. Shapiro, and J. M. Tranquada,
{\it Neutron scattering with a triple axis spectrometer} (Cambridge 
University Press, Cambridge, England, 2002).

\end{references}
\end{document}